\documentclass{article}

\usepackage{PRIMEarxiv}

\usepackage[utf8]{inputenc} 
\usepackage[T1]{fontenc} 
\usepackage{hyperref} 
\usepackage{url}  
\usepackage{booktabs} 
\usepackage{amsfonts} 
\usepackage{nicefrac} 
\usepackage{microtype} 
\usepackage{lipsum}
\usepackage{fancyhdr} 
\usepackage{graphicx} 
\graphicspath{{media/}} 

\title{Generative AI-based closed-loop fMRI system
}

\author{
 Mikihiro Kasahara \\
 Computational Neuroscience Labs \\
 ATR Institute International \\
 Kyoto\\
 \texttt{mi-kasahara@atr.jp} \\
 \And
 Taiki Oka \\
 Computational Neuroscience Labs \\
 ATR Institute International \\
 Kyoto\\
 \texttt{morioka\_sinri@atr.jp} \\
 \And
 Vincent Taschereau-Dumouchel \\
 Department of psychiatry and addictology \\
 Université de Montréal \\
 Montreal\\
 \texttt{vincent.taschereau-dumouchel@umontreal.ca} \\
 \And
 Mitsuo Kawato \\
 Computational Neuroscience Labs \\
 ATR Institute International \\
 Kyoto\\
 \texttt{kawato@atr.jp} \\
 \And
 Hiroki Takakura \\
 Center for Strategic Cyber Resilience \\Research and Development \\
 National Institute of Informatics \\
 Tokyo\\
 \texttt{takakura@nii.ac.jp} \\
 \And
 Aurelio Cortese \\
 Computational Neuroscience Labs \\
 ATR Institute International \\
 Kyoto\\
 \texttt{cortese.aurelio@gmail.com} \\
}

\begin{document}
\maketitle

\begin{abstract}
While generative AI is now widespread and useful in society, there are potential risks of misuse, e.g., unconsciously influencing cognitive processes or decision-making.
Although this causes a security problem in the cognitive domain, there has been no research about neural and computational mechanisms counteracting the impact of malicious generative AI in humans. We propose DecNefGAN, a novel framework that combines a generative adversarial system and a neural reinforcement model. More specifically, DecNefGAN bridges human and generative AI in a closed-loop system, with the AI creating stimuli that induce specific mental states, thus exerting external control over neural activity. The objective of the human is the opposite, to compete and reach an orthogonal mental state. This framework can contribute to elucidating how the human brain responds to and counteracts the potential influence of generative AI.

\end{abstract}


\section{Introduction}
The last decade can be described as the era of generative AI. 
Bloomberg Intelligence reported that the generative AI industry is expected to grow from 40 billion dollars in 2023 to 1.32 trillion dollars in 2032 [1].
Companies are beginning to use generative AI in all kinds of applications, from supporting the creation of presentations to generating ideas for projects and as a writing tool.
The rapid development of generative AI has also started directly affecting our daily lives.
In August 2023, the International Labour Organization reported that generative AI is unlikely to destroy human employment but rather lead to potential large-scale changes in the quality of work, particularly work intensity and automation [2].

In neuroscience research, generative AI has also been hugely influential. Recent studies successfully reconstructed natural images from recorded brain activity through Deep Neural Networks (DNN) [3-7].
This reconstruction framework can help visualize subjective internal representations, even accessing the content of visual illusions [8].
In parallel, a different line of research leveraged generative AI to generate visual stimuli that enhance targeted, specific internal states in the brain [9-10].

However, generative AI also poses potential risks. These risks are already observed in some domains, for example, there are generative AIs intended for actual abuse (e.g., WormGPT, FraudGPT [11]), which are used to assist in cyberattacks.
These attack tools are often devised by breaking system constraints, such as by entering special prompts in normal applications (e.g., ChatGPT, DALL-E2), and this jailbreaking can circumvent normal safeguards [12-14].
Currently, it is still unknown if such malicious use of generative AI could also breach the cognitive domain. By impacting our decision-making and cognitive processes in ways we may not be aware of, generative AI could potentially represent a threat.
Microsoft Copilot is an everyday example of generative AI that can potentially become a threat in our daily lives.
When performing an action such as a search, the generative AI works in the background, seemingly only optimizing results for the user.
However, potential risks exist as it can manipulate content with malicious intent. In such a scenario, this generative AI could fabricate search results and display, for instance, images intended to manipulate the user's psychological states, for instance, by evoking anger, fear or even specific beliefs.
Thus, by interacting with humans using specific stimuli, generative AI can potentially modulate human cognition and possibly even the internal representations of the human brain [10].
These possibilities should be considered as potential security risks for humans.
As described above, generative AI has already been used to activate specific representations in the brain. By exploiting this framework, malicious actors could make it possible to control internal representations in the human brain from the outside.
This might enable new but severe breaches of individuals' right to human integrity.

Thus, as AI becomes more widespread, it now appears important to understand how humans interact with generative AI [15]. In particular, we believe that it is crucial to study how the brain responds and adapts to generative AI and its hyper-realistic content. This new emerging field of research will likely develop into an important research theme at the intersection of neuroscience, cognitive science, psychology, computer science, and cybersecurity. 

Under these circumstances, there has been very little research yet bridging cybersecurity and human cognitive neuroscience in the context of generative AI. In particular, and to the best of our knowledge, there has been no research about the threat of generative AI to brain/cognitive functions, how humans are affected by malicious generative AI, and how humans may counteract generative AI.

To address these issues, we propose a new, neuroscience-driven framework that combines conventional neuroimaging and neural decoding with stimuli generation methods based on generative AI in a closed-loop fashion. This framework builds upon previous work in real-time neuroimaging [3-8] and iterative stimulus generation that maximally activates a target neural population [9-10]. Neural reinforcement allows one to strengthen brain activity and its associated behaviours with rewards or establish a subconscious link between particular brain patterns and target stimuli [16].
A key feature is that rather than measuring global brain activation in a given region, this method employs Multivoxel Pattern Analysis (MVPA) to decode nuanced brain patterns, meaning one can access specific information content from brain activity patterns.
Our recent works using neural reinforcement successfully showed one can guide the human brain's internal representations to a target state [16-20]. 
Over multiple studies, we have shown efficacy in altering cognitive functions, intervening to reduce fear memories, transforming confidence judgments, and manipulating preferences for faces without the participant's intent or awareness [17-20].

Here, we propose a novel framework by combining neural reinforcement interventions and generative AI. Through this architecture, we can construct an "as if Generative Adversarial Networks (GANs) [21] loop" setup of humans and generative AI. In short, the participant and AI compete against each other to achieve specific yet orthogonal goals at the neural level. Through this structure, future experiments will be able to examine how humans can counteract the influence of generative AI and what the underlying neural and computational mechanisms supporting this flexibility are.

\section{New closed-loop fMRI framework of human and generative AI}

\subsection{Overall Architecture}

The setup couples a human participant and generative AI in a closed loop while the participant's brain activity is measured with functional Magnetic Resonance Imaging (fMRI) (see Fig. 1) [22][23]. Here, we discuss the experimental architecture and how humans can counteract manipulations by malicious generative AI, leading to elucidating the neural and computational nature of an array of human-AI interactions.

First, the generative AI has two components, a decoder and a generator. The decoder receives patterns of brain activity as input, reading out the participants' current mental state. Based on the decoded information, the generator in turn produces new content (e.g., an image) to strengthen that mental state further. The new recorded brain activity is then immediately fed back to the decoder for the next trial.
By iterating over this process, the generative AI can converge onto stimuli that maximally activate an individual's internal brain representations [9-10].

\begin{figure}[h]
\begin{center}

\includegraphics[scale=0.50]{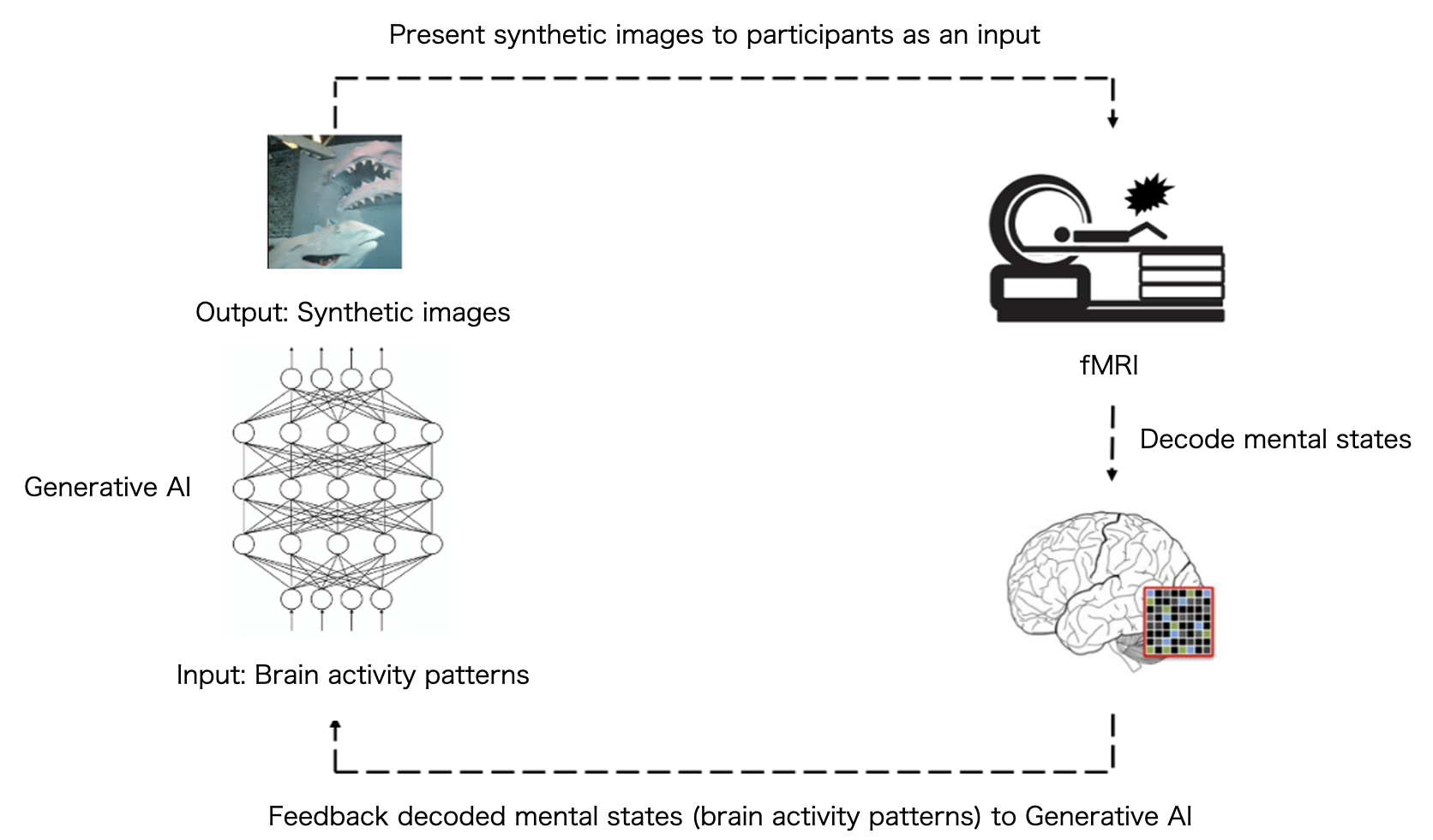}
\caption{Architecture of DecNefGAN (Decoded Neurofeedback based Generative Adversarial Networks of human and generative AI).
}
\end{center}
\end{figure}

\subsection{Generative AI architecture for specific mental status and human counteraction}
Before the procedure can be conducted, one needs to: (1) identify the specific/individualised mental states' category for the participant, (2) decode the specific mental states, and (3) generate AI-driven images based on the decoded mental state. These processes are then incorporated into the closed-loop neural reinforcement architecture.

In (1), the initial process identifies specific mental states in individual participants. 
Given the unique ways in which individuals may perceive the same world, people can interpret the same situation quite differently. This divergence arises from distinct internal representations within the human brain [24][25].
Such individual differences in internal representations imply that images to maximize specific mental states will vary from person to person. Therefore, it is crucial to tailor the approach to each individual's subjective experience in order to generate meaningful synthetic images.
To achieve this, we employ the Contrastive Language-Image Pre-training (CLIP) model, which uses image features and participants' self-reported ratings.
CLIP, trained on millions of annotated images available on the internet, globally describes the semantic concepts present in our world [26]. CLIP's potential semantic space may serve as the gold standard for quantifying an individual's subjective internal representations.
Concretely, participants are first presented with images and asked to subjectively assess the extent to which specific mental states are evoked. Subsequently, participants' reports are mapped to the CLIP model, yielding CLIP embeddings that consist of image features and ratings (see Fig. 2).
This enables the model to learn mappings between a participant's unique individual responses and image latent features, thus preparing the model to maximize individual mental states rather than considering a generic one across participants that may eventually not fit any individual well.

The second process is a "decoder construction" session, in which we train machine learning decoders that capture the brain activity patterns associated with the target mental state(s) of interest.
Here, we suggest using fMRI to collect brain activity data to be used to train the decoders; however, other modalities would probably be good too (e.g., MEG, EEG). 
Importantly, the decoder can be trained on binary labels (e.g., predicting one of two mental categories) or instead on the "degree" of a particular mental state (e.g., ratings on a scale).
By correlating the output degree of the decoded mental states with the CLIP model, we can align the strength of specific mental states decoded from brain activity with image features. This information will be most useful for the following step.

The third process concerns generating a synthetic image based on the level of the decoded mental state. Here, a generative AI can produce images using tools such as stable unCLIP [27], which is designed to invert the embeddings of images, allowing for the generation of synthetic images \textit{conditioned} on CLIP embeddings. In our framework, we input the strength of a participant's decoded mental state as ratings to the CLIP model. We can therefore obtain image embeddings and generate synthetic images reflecting the intensity of the targeted mental state.
Through this process, the generative AI can produce images that (at least theoretically) should maximally enhance an individual's target mental state.

\begin{figure}[h]
\begin{center}

\includegraphics[scale=0.54]{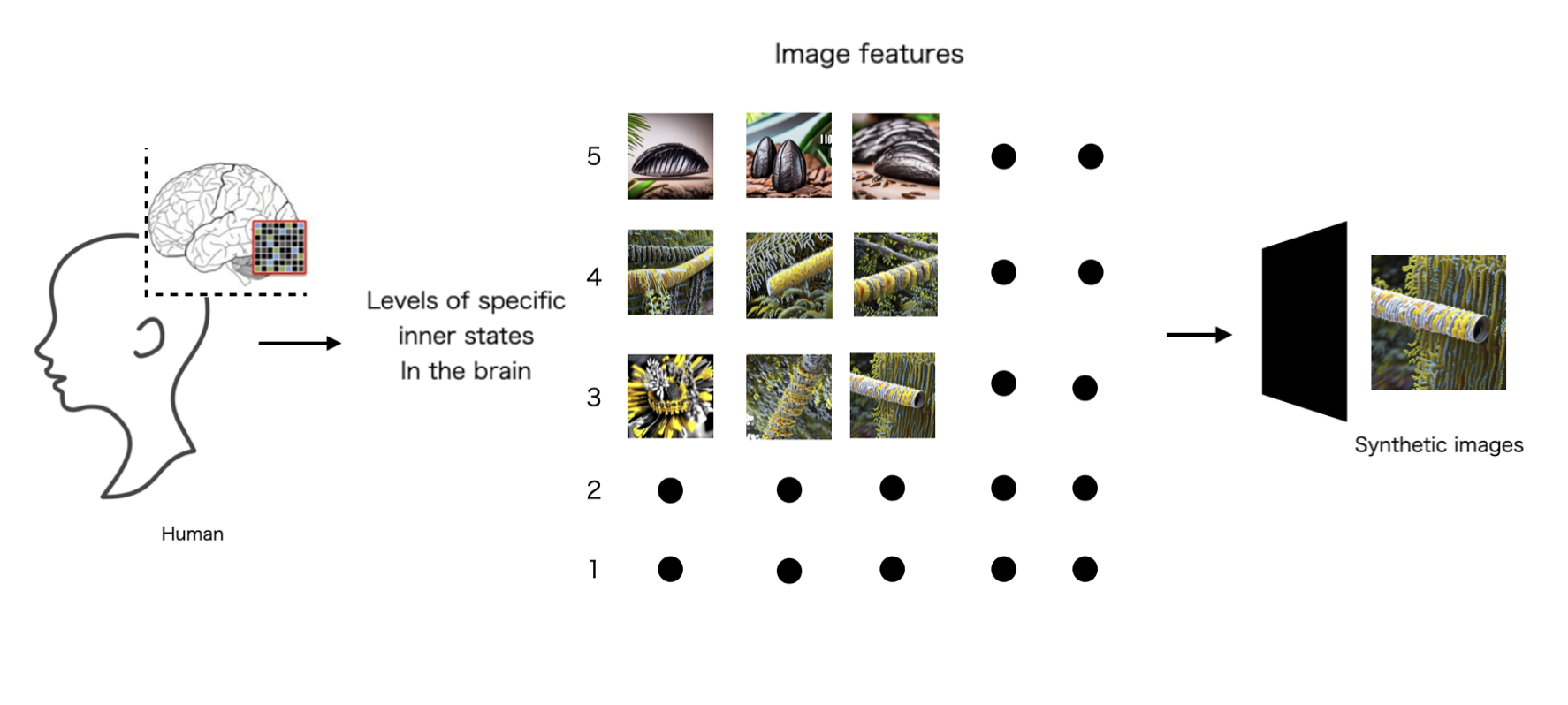}
\caption{Generative process based on decoding of latent brain states. Decoding the strength of a mental state from the brain, then associating the strength of the decoded mental state with pre-collected ratings from the participants. 
The mapped strength of the identified mental state is then correlated with image features in the latent representations of CLIP. 
From these latent spaces, generating images corresponding to levels of a specific mental state becomes possible.}

\end{center}
\end{figure} 

Finally, the constructed generative AI is incorporated into a closed-loop neural reinforcement framework. As mentioned, although AI can output images to maximize specific mental states, the participant can actively alter their own neural activity in response to the output from generative AI. Specifically, within our framework, when generative AI decodes a particular mental state, a human voluntarily seeks to alter their brain activity patterns so that generative AI cannot precisely decode that mental state. This proactive adjustment by a human aims to make it challenging for generative AI to produce visual stimuli that maximize a specific mental state. However, leveraging the participant's altered neural activity, generative AI may \textit{regenerate} images that can still induce the desired mental state. Thus, this framework forms a closed, functioning "as if Generative Adversarial Network (GAN)" loop between generative AI and humans.

In sum, by combining the concepts of closed-loop neurofeedback, generative AI, and GAN, we can explore the mechanisms at play when a human and generative AI engage in mutual opposition. Using this framework, investigators will be able to understand how a human can operate against generative AI and assess the extent to which a human can withstand generative AI's powerful iterative approach. In parallel, we can efficiently explore how generative AI adapts its behaviour when encountering resistance from a human.

A previous study has suggested that individuals can effectively protect their internal representations from external read-out by intentionally resisting decoding internal representations in the brain [28]. However, to comprehend the inherent risks posed by fundamentally malicious generative AI and understand the mechanisms of both human and generative AI, there is a need to elucidate the mechanisms within the confrontational dynamics of a closed-loop interaction [29].

Through DecNefGAN, as a human confronts and counters malicious generative AI, it can reveal the potential for resistance, as well as the adaptive strategies that may be used by generative AI to achieve its preset goals. This understanding will form the basis for implementing appropriate safeguards in the cognitive and cybersecurity domains. Taken together, our proposed DecNefGAN stands as a novel framework that leverages neuroscience to address adversarial strategies against generative AI and also presents an opportunity to enhance cybersecurity measures.

\subsection{Potential issues and solutions}
Two potential issues of DecNefGAN need to be considered: the machine learning decoding algorithm and the validity of the methods for representing subjective experience.
Firstly, decoding algorithms are essential to predict certain mental states appropriately.
Our framework is based on fMRI scanning of brain activity and generates visual stimuli corresponding to the level of the decoded mental state. If the decoding result is inaccurate, the generated synthetic images will not properly reflect the participant's internal brain representations, and the process may diverge towards unpredicted directions.
Our previous work has shown that we can read out fear states [30] from neural activity using our proposed CLIP-based decoder architecture; however, further research is required to determine whether decoding can continue stably in other scenarios.

The second concern is whether subjective experiences can be objectively quantified. Existing methods for measuring subjective experiences have been sometimes criticized for lacking consistency across different tools.
In addressing this issue, the CLIP model has been demonstrated to represent not only emotions like surprise or fear in semantic space but also higher-level semantic concepts such as geographical, political, and religious elements [31]. This feature contributes to quantifying results and enhancing consistency in result interpretation.
Our previous research successfully used CLIP to predict fear evaluations reported by specific participants [30]. We believe that this approach suggests the capability of more closely representing subjective experiences.

\subsection{Applications}
DecNefGAN can be applied to a variety of purposes.
Through the aforementioned series of processes, we can contribute to elucidating how the human brain/cognitive function counteracts the influence of generative AI and what mechanisms are involved in this resistance, possibly revealing important mechanisms for subjective experience to emerge [32][33][34].
This approach has the potential to lead to safeguards against human-targeted attacks by malicious generative AI, contributing to security in the realm of human cognitive domains (e.g. there is the potential to prevent actions that could cause harm to our society stemming from human manipulation by malicious generative AI for mind control).
The process in our proposed framework, where generative AI repeatedly presents stimuli to humans, bears similarity to PTSD therapeutic methods that mitigate fear by repeatedly presenting fear stimuli [35]. Leveraging this similarity, there is potential for generative AI to play a role in PTSD treatment, as participants learn to associate new neutral responses with the fear stimuli presented by the AI.
It may also be applied to learning purposes, for example, to facilitate the rapid acquisition of new skills.

\section{Ethical challenges}
Although we argue that our proposed framework helps explore neurocognitive mechanisms that counteract generative AI, it also faces significant ethical issues.

First, DecNefGAN is based on neural reinforcement (i.e., decoded neurofeedback). Although research to date has shown that neural reinforcement strengthens and influences internal neural representations below consciousness in a targeted and specific manner [16-20], such neuromodulation techniques may also change brain activity in unintended directions.
In particular, DecNefGAN personalizes generative stimuli according to the participant's internal state, and by doing so, the generative AI could powerfully affect the individual's brain/cognition. For instance, if generative AI produces highly fearful categories to maximize a participant's fear, this could cause the participant to experience intense psychological distress. 

As a countermeasure to this problem, one can use an approach that minimizes the conscious impact.
To reduce the participants' stress, our past study has guided fear responses towards reducing fear when rewarding brain activity patterns associated with the ventral visual stream resemble those of the fear category. As a result, we successfully mitigated defensive responses to the fearful category without consciously accessing the feared category [20][36].
Our past research has shown the ability to induce activity patterns of a small number of objects without actually evoking fear, revealing a dissociation between mental state representation and consciousness.
This method avoids showing the category of the feared object directly and might be a solution to the ethical problem of consciously presenting the fear stimuli.
However, other ethical issues may arise when using generative AI and neural reinforcement technology (e.g., altering brain activity patterns using generative AI could lead to mind control).
Traditionally, such issues have been assessed based on individual studies' ethical and safety reviews rooted in the Declaration of Helsinki [37]. However, ethical guidelines for determining the appropriateness of research designs concerning emerging technologies, like generative AI, are yet to be firmly established.
In recent years, there has been a global movement to develop a framework for neuroethics (e.g., IEEE, UNESCO, [38-39]), and some have begun to contemplate the ethics of generative AI within the context of scientific writing and research [40-41]. 
Nevertheless, ethical research frameworks that combine neuroscience and generative AI remain incomplete. Further theoretical research is required to elucidate and solve ethical issues of generative AI combined with neuroscience approaches that have the power to change brain, cognitive or behavioural states.

\section{Concluding remarks}
In this opinion paper, we introduced DecNefGAN, a novel framework integrating generative AI and neural decoding with closed-loop fMRI. 
We discuss DecNefGAN and its application to elucidate how humans can counteract potentially malicious, generative AI manipulations.
We suggest this framework can be extended from security in the human cognitive domain to, more broadly, the field of mental treatments in psychiatry and clinical psychology, contributing to the emergence of a new era in human-AI interaction.
Future works include verifying the effectiveness of DecNefGAN through behavioural experiments and aim to contribute to establishing a research field related to the security of cognitive function/brain.

\section*{Acknowledgements}
This research was supported by JSPS KAKENHI Grant Number JP22H05156; Brain-AI Hybrid ERATO grant (JPMJER1801) from the Japan Science and Technology Agency; Project of Cyber Security Establishment with Inter-University Cooperation from Ministry of Education, Culture, Sports, Science and Technology; KDDI collaborative research contract; Innovative Science and Technology Initiative for Security Grant Number JPJ004596, ATLA, Japan.

\section*{References}
[1] Bloomberg Intelligence. (2023) Generative AI to Become a \$1.3 Trillion Market by 2032, Research Finds. Accessed on October 30th, https://www.bloomberg.com/company/press/generative-ai-to-become-a-1-3-trillion-market-by-2032-research-finds/.

[2] Paweł G. et al. (2023) AI and jobs: A global
analysis of potential effects on job
quantity and quality. ILO Working Paper 96.

[3] T. Horikawa et al. (2017) Generic decoding of seen and imagined objects using hierarchical visual features. Nat. Commun. 8, 15037.

[4] K. Seeliger et al. (2018) Generative adversarial networks for reconstructing natural images from brain activity. Neuroimage 181, 775–785.

[5] G. Shen et al. (2019) Deep image reconstruction from human brain activity. PLOS Comput. Biol. 15, 1006633.

[6] G. Shen et al. (2019) End-to-end deep image reconstruction from human brain activity. Front. Comput. Neurosci. 13, 21.

[7] T. Horikawa et al. (2022) Attention modulates neural representation to render reconstructions according to subjective appearance. Commun. Biol. 5, 34.

[8] Fan L. et al. (2023) Reconstructing visual illusory experiences from human brain. activity. https://www.science.org/doi/full/10.1126/sciadv.adj3906

[9] Pouya B. et al. (2019) Neural population control via deep image synthesis.Science 364.

[10] Carlos R. et al. (2019) Evolving Images for Visual Neurons Using a Deep Generative Network Reveals Coding Principles and Neuronal Preferences. Cell, 177, 999-1009.

[11] Maximilian M. et al. (2023) Use of LLMs for Illicit Purposes: Threats, Prevention Measures, and Vulnerabilities. arXiv arXiv:2308.12833.

[12] Yuchen Y. et al. (2023) SneakyPrompt: Jailbreaking Text-to-image Generative Models. arXiv arXiv:2305.12082.

[13] Andy Z. et al. (2023) Universal and Transferable Adversarial Attacks on Aligned Language Models. arXiv arXiv:2307.15043.

[14] Jiahao Y. et al. (2023) GPTFUZZER: Red Teaming Large Language Models with Auto-Generated Jailbreak Prompts. arXiv arXiv:2309.10253.

[15] Shi et al. (2023) An HCI-Centric Survey and Taxonomy of Human-Generative-AI Interactions. arXiv arXiv:2310.07127v1.

[16] Shibata K. et al. (2011) Perceptual Learning Incepted by Decoded fMRI Neurofeedback Without Stimulus Presentation. Science 334,1413-1415.

[17] Koizumi, A. et al. (2017) Fear reduction without fear through reinforcement of neural activity that bypasses conscious exposure. Nat Hum Behav 1, 0006.

[18] Aurelio C. et al. (2016) Multivoxel neurofeedback selectively modulates confidence without changing perceptual performance. Nature Communications. Nat Commun 7, 13669. 

[19] Kazuhisa S. et al. (2016) Differential activation patterns in the same brain region led to opposite emotional states. PLoS Biol 14, 9.

[20] Vincent T. et al. (2018) Towards an unconscious neural reinforcement intervention for common fears. PNAS 115, 3470-3475.

[21] Goodfellow et al. (2014) Generative Adversarial Networks. arXiv arXiv:1406.2661.

[22] Vincent Taschereau-Dumouchel, Mathieu Roy. (2020) Could Brain Decoding Machines Change Our Minds? Trends in Cognitive Sciences, 24(11). 

[23] Vincent Taschereau-Dumouchel, Cody A. Cushing, and Hakwan Lau. (2022) Real-Time Functional MRI in the Treatment of Mental Health Disorders. Annual Review of Clinical Psychology 18, 125-154.

[24] Finn, Emily, Xilin Shen, Dustin Scheinost, Monica Rosenberg, Jessica Huang, Marvin Chun, Xenophon Papademetris, and R. Constable. 2015. “Functional Connectome Fingerprinting: Identifying Individuals Using Patterns of Brain Connectivity.” Nature Neuroscience 18 (11): 1664–71.

[25] Finn, Emily S., Enrico Glerean, Arman Y. Khojandi, Dylan Nielson, Peter J. Molfese, Daniel A. Handwerker, and Peter A. Bandettini. 2020. “Idiosynchrony: From Shared Responses to Individual Differences during Naturalistic Neuroimaging.” NeuroImage 215 (July): 116828.

[26] Alec Radford, Jong Wook Kim, Chris Hallacy, Aditya Ramesh, Gabriel Goh, Sandhini Agarwal, Girish Sastry, Amanda Askell, Pamela Mishkin, Jack Clark, Gretchen Krueger, Ilya Sutskever. (2021) Learning Transferable Visual Models From Natural Language Supervision. arXiv arXiv:2103.00020.

[27] Aditya R. et al. (2022) Hierarchical Text-Conditional Image Generation with CLIP Latents. arXiv arXiv:2204.06125.

[28] Tang, J. et al. (2023) Semantic reconstruction of continuous language from non-invasive brain recordings. Nat Neurosci 26, 858–866.


[29] Sundar et al. (2020) Rise of Machine Agency: A Framework for Studying the Psychology of Human–AI Interaction (HAII). Journal of Computer-Mediated Communication 25, 74-88. 

[30] Taschereau-Dumouchel et al. (2023) Interaction Between the Prefrontal and Visual Cortices Supports Subjective Fear. bioRxiv 2023.10.23.562918.

[31] Gohnick et al. (2021) Multimodal Neurons in Artificial Neural Networks. Distill 10.23915/distill.00030. 

[32] Lau, H. (2019). Consciousness, Metacognition, \& Perceptual Reality Monitoring. PsyArXiv. 10.31234/osf.io/ckbyf.

[33] Samuel J. Gershman. (2019). The Generative Adversarial Brain. Front. Artif. Intell., 18. 

[34] Cody A Cushing, Alexei J Dawes, Stefan G Hofmann, Hakwan Lau, Joseph E LeDoux, Vincent Taschereau-Dumouchel. (2023). A generative adversarial model of intrusive imagery in the human brain. PNAS Nexus, 2(1).

[35] MG Craske, et al., (2008) Optimizing inhibitory learning during exposure therapy. Behav Res Ther 46, 5–27.

[36] Cody A. Cushing, Hakwan Lau, Mitsuo Kawato, Michelle G. Craske, Vincent Taschereau-Dumouchel. (2023) A pre-registered decoded neurofeedback intervention for specific phobias. medRxiv 2023.04.25.23289107.

[37] World Medical Association (2013). Declaration of Helsinki: Ethical Principles for Medical Research Involving Human Subjects. JAMA. 310, 2191–2194.

[38] IEEE Neuroethics Framework. (2023) Addressing the Ethical, Legal, Social, and Cultural Implications of Neurotechnology. Accessed on November 25th. https://brain.ieee.org/publications/ieee-neuroethics-framework/

[39] UNESCO. (2023) Ethics of Neurotechnology: UNESCO, leaders and top experts call for solid governance. Accessed on November 13th. https://www.unesco.org/en/articles/ethics-neurotechnology-unesco-leaders-and-top-experts-call-solid-governance

[40] Giovanni E. Cacciamani, Gary S. Collins, Inderbir S. Gill. (2023) ChatGPT: standard reporting guidelines for responsible use. Accessed on December 27th.
https://www.nature.com/articles/d41586-023-01853-w

[41] Eva A. M. van Dis, Johan Bollen, Willem Zuidema, Robert van Rooij, Claudi L. Bockting. (2023) ChatGPT: five priorities for research. Accessed on December 27th. 
https://www.nature.com/articles/d41586-023-00288-7

\bibliographystyle{unsrt} 
\end{document}